\documentstyle[12pt,aasms] {article}
\tighten
\begin{document}

\title{The MACHO Data Pipeline}

\author{T. S. Axelrod}
\affil{Mount Stromlo and Siding Springs Observatory, Canberra, ACT 2611, Australia}
\affil{Lawrence Livermore Laboratory, Livermore, CA 94550}
\author{R. A. Allsman}
\affil{Supercomputing Facility, Australian National University, Canberra, ACT 0200, Australia }
\author{P. J. Quinn}
\affil{European Southern Obervatory, Karl-Swarzchild-Str. 2, D-85748 Garching bei Munchen, Germany}
\author{D. P. Bennett}
\affil{Center for Particle Astrophysics, University of California, Berkeley, CA 94720}
\author{K. C. Freeman, B.A. Peterson, A. W. Rodgers}
\affil{Mount Stromlo and Siding Springs Observatory, Canberra, ACT 2611, Australia}
\author{C. Alcock, K. H. Cook}
\affil{Lawrence Livermore Laboratory, Livermore, CA 94550}
\author{K. Griest}
\affil{Department of Physics, University of California, San Diego, CA 92093}
\author{S. L. Marshall, M. R. Pratt, C. W. Stubbs}
\affil{Departments of Astronomy and Physics, University of Washington, Seattle, WA 98195}
\and
\author{W. Sutherland}
\affil{Department of Physics, Oxford University, Oxford OX1 3RH, U.K.}
\
\affil{Electronic mail: tsa@mso.anu.edu.au, robyn@macho.anu.edu.au, pjq@eso.org,
bennett@igpp.llnl.gov, kcf@mso.anu.edu.au, peterson@mso.anu.edu.au, alex@mso.anu.edu.au,
alcock@igpp.llnl.gov, kcook@igpp.llnl.gov, griest@astrophys.ucsd.edu,
stuart@astro.washington.edu, mrp@astro.washington.edu, stubbs@astro.washington.edu, 
sutherland@physics.oxford.ac.uk}
\begin{abstract}

The MACHO experiment \markcite{(Alcock 1992, 1993)} is searching for dark matter in
the halo of the Galaxy by monitoring more than 20 million stars in the
LMC and Galactic bulge for gravitational microlensing events.  The
hardware consists of a 50 inch telescope \markcite{(Hart 1995)}, a two-color 32
megapixel ccd camera \markcite{(Stubbs 1993)} and a network of computers.  On clear
nights the system generates up to 8 GB of raw data and 1 GB of
reduced data.  The computer system is responsible for all realtime
control tasks, for data reduction, and for storing all data associated
with each observation in a data base.  The subject of this paper is the software
system that handles these functions.  It is an integrated
system controlled by Petri nets that
consists of multiple processes communicating via mailboxes and a
bulletin board.  The system is highly automated, readily extensible,
and incorporates flexible error recovery capabilities.  It is implemented
with C++ in a Unix environment.
\end{abstract}

\section{SYSTEM CHARACTERISTICS}

	During the earliest discussions of the feasibility of the MACHO
project the team recognized that the software would be one of the two
most challenging system components (the other being the ccd camera).
Before the software design process began, we set ourselves a number of
goals for the system.  We discuss them first, as  a prelude to
the following sections, which cover the architecture and implementation
of the system.

\subsection{Functionality}

The scientific goals of the MACHO project \markcite{(Alcock 1992, 1993)} require that
the system operate every clear nighttime hour with high reliability.
The duration of the data collection phase was initially set as 4
years.  We recognized that the data must be processed, as
well as collected, in near real-time.  Otherwise, the project was
likely to die from data indigestion.  Additionally, there is significant
scientific gain from being able to detect microlensing events while they
are still in progress.  We decided at the outset that
these requirements could be met only by a system which encompassed in
an integrated framework all the separate tasks required for operation
of the experiment, and one which operated in a highly automated way
with minimal human intervention.  The main tasks required of the system
are:

	1. Scheduling of observations

	2. Control of the telescope

	3. Operation of the camera system

	4. Flatfielding of the ccd images

	5. Archiving the images to tape

	6. Preparing the images for photometry

	7. Photometry

	8. Storing all information associated with each observation in a 
	data base

	9. Interacting with the human operator

Since the Macho telescope is dedicated to a single experiment, this
software can run continously, and can be highly specialized.

Typical exposure times are 150 to 600 sec, with times as short as a few sec 
sometimes being employed.  This time is short enough that the time 
required to move the telescope and to read out the ccd camera 
system is significant.  To maintain the maximum possible rate of 
observations, it is therefore crucial that the various system tasks 
be overlapped as much as possible (e.g move the telescope at the 
same time as reading out the camera;  flat field the preceding image 
while exposing the next).

Additionally, given the very large number of pixels in the camera system
and the high star densities that are typical of our fields, the photometry
process is very cpu intensive.  The use of multiple cpu's is absolutely
essential to process data at the rate it is taken.

\subsection{Hardware Configuration}

	The hardware configuration, and the underlying operating system
software, has evolved significantly since the beginning of the
project.  At the outset, the main computer was a Solbourne 5/804, a 4
processor Unix system running a version of SunOS 4.1.  Today, the main
computer is a Sun SparcServer 1000 with 8 cpu's and 0.5GB of memory,
running Solaris 2.4, supported by a secondary SparcServer 1000 with 4
cpu's.  Online disk storage has grown from an initial 2GB to
approximately 150GB today, and is still growing rapidly.  The current
hardware configuration is shown in Fig. 1.

\subsection{Need for Robustness}

	Given the complexity of the system, a high priority was given 
to achieving robustness in an environment where a wide variety of 
error conditions can occur, and occasional software ``crashes'' are 
inevitable.  Several sub-goals were identified that together would 
contribute to the desired robustness:

\subsubsection{Persistent State}

The system inevitably has a large amount of ``state'' associated with 
it.  Examples of state variables include the current observation 
number, the status of an observation in the observe cycle, and the 
information needed to identify and communicate with the processes 
that make up the system.  We recognized early that making the state 
persistent would be a big contributor to robustness.  Our meaning of 
persistence is simply that information is preserved when a process 
dies, rather than disappearing with the process.  Persistent state 
allows a replacement process to be started and take its place in the 
system without special action to tell it what it needs to know.

\subsubsection{Central Heartbeat Monitoring}

A failure mode which is particularly troublesome in systems of this 
nature occurs when a process enters a ``hung'' state in which it 
continues to exist but is no longer doing its job.  When this occurs, 
gridlock often ensues as other processes wait for the hung process to 
do something useful.  A human operator, faced with such gridlock, 
seldom has the information necessary to determine the cause and 
generally has no option but to restart the entire system.
We decided that many such failures could be recognized if every 
member process was endowed with an externally visible ``heartbeat''.  
One process could then be assigned the task of monitoring the 
heartbeats of all component processes.  Disappearance of the 
heartbeat would occur not only when the process itself disappears 
(i.e. ``crashes''), but also in many cases where the process ``hangs''.  
Heartbeat existence is useful both for automated recovery 
procedures and for the human operator, who can now see which 
process is hung and take action less drastic than a global restart.

\subsubsection{Capability to post and respond to alarms}

When an error condition is detected it is important to have a variety 
of options for responding to it.  The response can range from an 
automatic ``reflex'' on the part of the process which recognizes the 
error, such as ``retry the operation'', to one which alerts the human 
operator and gives him the opportunity to respond in an informed 
way.  We decided that it was crucial to have a universal mechanism 
for posting alarm conditions and allowing a variety of responses.
 
\section{SYSTEM ARCHITECTURE}

\subsection{Observation Cycle}

The architecture we chose has as its central feature the observation
cycle, and the mapping of that cycle to communicating Unix processes.
The observation cycle, shown schematically in Fig. 2,  is a fixed set
of operations that are performed in the same order each time a new
observation is taken.  Additionally, we make use of the same
observation cycle for ``reobserving'', which is our term for performing
the data reduction steps on a previously taken observation.  This
situation comes up frequently, usually because photometry was not
performed at the time of exposure, or because some problem requires it
to be repeated.  During reobserving, all stages of the observing
pipeline are active, but some of them (e.g. Telescope and Camera) have
little to do.

A key concept in the observation cycle is the observation descriptor.
Observation descriptors are used by all phases of the observation cycle
and completely describe the status of an observation and the parameters
associated with it as it moves through the pipeline.  When an
observation is completed, its descriptor is saved permanently in a
data base.  Table 1 shows the contents of a representative observation descriptor,
as printed out by a viewer utility.
The observation descriptor has proved to be an important unifying data
structure, and it has expanded several times in the course of the
project when it became clear that it would be valuable to include more
data (for example, the inclusion of telescope temperature data
was a recent addition).  Its uniform use throughout all phases of the
observation cycle has considerably simplified both the initial
implementation and subsequent changes to the observation cycle.  It has
also made it easy for the progress of an observation to be monitored
from the user interface (UI).

Here is a brief description of each operation in the observation cycle:

\subsubsection{Scheduler}

The scheduler gathers the information needed to specify the next observation,
builds its observation descriptor, and a creates a number of associated 
directories.  The required information includes the field number (if it is
a prespecified Macho field, otherwise the pointing coordinates), the exposure
time, and whether or not photometry is to be performed.  The information
can come either directly from the observer through the user interface (UI),
from an ascii file, or from an external source that places observation
requests on the scheduler's work queue.  Which source is active is determined
by the observer through the UI.

\subsubsection{Slotter}

Every new observation requires a number of records in the data base structure
to be created and initialized for it.  This is the main role of the slotter.

\subsubsection{Telescope}

The telescope drive is controlled by a dedicated VAX/VMS computer that is not
itself part of the Macho system \markcite{(Hart 1995)}.  The main function of the telescope
process is to issue the proper commands to the telescope computer so that it
points and tracks at the coordinates required for the observation.  
The telescope coordinates are continuously monitored to ensure that the 
correct pointing is in fact achieved.  The telescope process maintains on
the bulletin board (see Section 3.1.3) a current descriptor of the telescope state
for use by other processes.  Additionally, the telescope process has the
capability to automatically set the telescope focus.  This feature is not
currently in use, because it has proved difficult to create a satisfactory
focus model - but we think we are now close.

\subsubsection{Camera}

The camera process issues commands to the camera controller (a micro
computer running Forth) to properly initialize the camera, take the
exposure, and read out the image \markcite{(Stubbs 1993)}.  It then
makes use of a separate server process which flatfields the image,
and corrects for crosstalk between ccd channels \markcite{(Stubbs 1995)}, 
and finally stores the resulting images on disk as FITS files.

\subsubsection{Image Archiver}

The archiver writes the flatfielded images to Exabyte tape for archival 
storage.  It is a simple procedure, but must nonetheless contend with a
variety of error conditions (I/O errors, end of tape, etc) in a robust
way.

\subsubsection{Slicer}

All data reductions are performed not on whole ccd images, which at 2K x 2K are
too large for convenience, but on ``chunks'', which are roughly 500 pixels square.
These chunks are overlapping, so that unnecesssary edge effects are not 
introduced.  The slicer is responsible for creating the chunk images, creating
for each chunk a data structure that binds together all data necessary for processing it,
and mailing it off to the photometry work queue.

\subsubsection{Photometry}

The photometry process is responsible for running the SoDoPhoT \markcite{(Bennett 1995)} photometry
package on each of the image chunks.  A data structure is created for each chunk
that binds together all resulting photometry data for the chunk.  This data structure
is then mailed off to the collector work queue.  A typical chunk contains 5000 
stars and takes 20 sec to process.

\subsubsection{Collector}

The collector is the final stage of the observation cycle.  The photometry data for
each chunk is loaded into the photometry data base, where it is then available for
later analysis.  Database loading rates are typically 5000 stars/sec, so that
a chunk requires roughly one sec to process.

\subsubsection{User Interface}

The principal functions of the UI are to :

	1. Allow the observer to initiate, pause and stop the observation cycle,
	   and specify parameters that affect it, such as exposure times.

	2. Present process status (heartbeats) in a graphical form.

	3. Present alarms for action by the observer.

	4. Allow the observer to assess the progress of any observation through
	   the pipeline and intervene if necessary.

\subsubsection{Controller}

The controller process coordinates the overall functioning of the system.
It handles the startup of all component processes, monitors their health,
and restarts failed processes
when required.  It also contains the central sequencing logic for the observation
cycle and the maintenance cycle (discussed below).

\subsection{Parallel Architecture}

The component operations of the observation cycle form a data 
pipeline, with the output of one operation being the input to the next.  
As with any pipeline, under conditions of continuous operation 
higher throughput may be achieved by assigning each pipeline stage 
to its own hardware, so that it operates in parallel with the others.  

Because of the unchanging structure of the observation cycle, the ``macro''
nature of each of the pipeline stages, and the desirability of having
the stages handled by different processors, we decided to map each
stage to a Unix process.  In the case of the photometry stage, which is
by far the most compute intensive phase of the observation cycle, we
assign a group of processes (typically 4 or 5) to the function, further increasing
parallelism.  This could readily be done for other compute intensive
stages of the pipeline if their throughput ever becomes an issue.

The MACHO system makes explicit two levels of parallelism.  The first
level, as mentioned above, is that between the Unix processes that
cooperate to form the observation pipeline.  Within each of these
processes there is a second level of parallelism, which is present to
deal with the demands of handling asynchronous events such as the
arrival of messages and the expiration of timers.

The motivation for the process level, or coarse grain, parallelism is
principally the need to exploit multiple CPU's for increased
performance.  Secondarily, the multiple process structure is desirable
from a software engineering point of view, because the process
boundaries enforce a useful modularity on the system, and eliminate
many opportunities for unforseen interaction between components.

The principal motivation for the second level of parallelism, which we
will call the fine grain level, is quite different.  The problems of
creating reliable real-time control systems that handle multiple
classes of asynchronous events are well known \markcite{(Everett
1995)}.  Avoiding deadlocks and other forms of ``hangs'',  and ensuring
that all incoming events are processed in a fair way are all difficult
in a sequential program.  On the other hand, a programming method that
uses parallel threads of control can solve many of these problems,
especially if it is practical to spawn a new thread to process each
incoming event.

Secondarily, multiple threads have allowed major performance improvements
to be made in I/O intensive processes, such as the collector.

\subsection{Petri Nets: Unifying Coarse- and Fine-grained parallelism}

One innovation of the Macho system is that it ties together the coarse-
and fine-grain levels of parallelism into a single unified view of the
system.  This is valuable both for system understanding, and for
efficient implementation of the system.  This unified view is based on
Petri nets, which we now briefly discuss.

Petri nets are a well known methodology for designing and, less
commonly, for implementing parallel systems \markcite{(Murata 1989)}.
In its most basic form a Petri net consists of a set of places; a set
of transitions; and a set of tokens (Fig. 3).  The execution rules for
the net are:

	1. A transition ``fires'' when all of its input places are marked
	with tokens.

	2. When a transition ``fires'', it marks each of its output
	places with a token.

This structure allows a surprising variety of parallel, sequential, and hybrid
systems to be modeled.  However, in this simple form, a Petri net lacks a number
of features that are required in a real system implementation:

	1. There is no way of performing ``useful work'', e.g. performing a
	computation or writing a file.

	2. There is no conditional behavior

	3. There is no way of expressing timeout conditions.

Petri nets have been extended in many different ways to overcome these,
and other, difficulties.  Our extensions are inspired by \markcite{(DiStefano
1991)}, one of the few examples of using Petri nets for the actual
implementation of a real time system, and consist of the following:

	1. A transition can optionally have an associated function that
	is executed whenever the transition fires.  Tokens are allowed
	to carry optional pointers to arbitrary data which the
	transition function can utilize as input and output arguments.

	2. Conditional transitions are defined, which mark either a
	``true'' or a ``false'' output place, based on the return value of
	an associated function.

	3. Timeout transitions are defined, which always have two
	input, and two output places.  Marking one of the input places
	starts the timer, while marking the other stops it.  If the
	timer runs for longer than a preset timeout interval, the first
	output place is marked, while if the timer is stopped before
	the interval expires, the second output place is marked.   This
	definition is identical to that used in \markcite{(DiStefano 1991)}.

\subsection{Petri Nets that cross process boundaries}

Now we must take a step closer to implementation, and consider Petri
nets in the context of Unix processes executing in parallel.  It is
here that we are able to unify the coarse- and fine-grained levels of
parallelism.

It is natural to express each stage of the observation cycle as an
extended Petri net.  These nets are connected because the
pipeline stages must communicate.  If we have a mechanism to allow
tokens to cross process boundaries, we can then use this to connect the
nets, and thereby end up with a Petri net description of the entire
system that spans process boundaries in a transparent way.  A sketch of
this concept is shown in Fig. 4.

\subsection{Database}

The operation of the data pipeline requires a large amount of input 
data in addition to the images taken by the camera system.  This data
includes flat and dark images for the camera system, the definitions of
field centers, photometric reduction parameters, locations of fiducial stars, and
a great deal more.  Additionally, the pipeline generates a large amount of
output data for every observation.  The largest volume of output data
consists of images and star photometry, but there is much else, 
including log files, temperature data, and so on.

The data base system is designed to organize all this input and output
data in a manner that supports a variety of analysis tasks.  It is
essential that the data base preserve the history behind each observation:
full details of the environmental conditions when the exposure was taken,
the operations that were performed on the data, and the identity of the
software that performed them.

The ramifications of this requirement can be quite extensive.  As an 
example, several of the focalplane ccd's were recently replaced.  The
replacements have somewhat different orientations than their predecessors,
as well as different darks, flats, readout noise and crosstalk parameters.  
The data base must keep track of these changes, and make sure that the proper ccd
data is associated with each observation in the data base, and that this data
is transparently routed to any analysis programs that require it.

The data base defines a few fundamental geometrical data structures:

1. A Field is the nominal portion of the sky captured on the ccd
during an exposure.  Fields may overlap. The geometry of a field is 
considered inviolate once it is ingested into the Field data base. 
Fields are currently 43 arc minutes on a side, and frequently contain
500,000 stars or more.

2. A Chunk is a portion of ccd image, typically of order 500 pixels on
a side. Its corresponding position in the sky is known only to an
approximation.  Chunks overlap in order to avoid needless photometry
errors near chunk boundaries. The geometry of a chunk may change over
the life of the project.

3. A Domain is a bounded region in the sky which totally encompasses a
pre-defined set of fields. A Domain is divided into an evenly spaced
rectangular grid. Each grid region is known as a Tile.  A Tile is
typically 4 arc minutes on a side, and contains several thousand
stars.  The Domain geometry and its grid definition are considered
inviolate once they are ingested into the Domain data base.

4. A Template is a set of stars from a given Field which is fabricated
to completely cover a ccd Chunk. A Template may contain stars from
multiple Tiles.  The geometry of a Template is tied to the geometry of
a Chunk and may change over the life of the project. Also, the star set
may be modified without changing the geometry.

5. An Image is an aggregate output of 4 red- and 4 blue-band ccd's.
The ordering and color assignments of ccd's may change over the life of
the project.  

The relationship of these structures is shown in Fig. 5.

The data base system has two levels of organization.  The primary data base files contain
some particular record type spanning some range of time, for example, 
field boundary definitions or ccd characteristics. The second tier collects 
the timestamped primary data bases into a set containing the full span of time varying
characteristics of observations and their photometry.

The most important data base files are as follows:

	1. The Domain data base defines Domain boundaries and tiling size.  The tile size 
	is selected so each Tile contains an average of 5 thousand stars.
	
	2. The Field data base defines the Telescope pointing coordinates.
	
	3. The FieldChunk data base defines the template geometry. Since its
	data may change over the course of the project, it is tagged so that the
	version corresponding to any observation is automatically retrieved.
	
	4. A Star data base encompasses the coordinates of all fiducial and observed 
	stars within a single Tile's boundary.  
	
	5. The CCD data base defines the focal plane characteristics. Since its
	data may change over the course of the project, it is tagged.
	
	6. The CDF data base contains crosstalk, dark, and flatfield information for 
	the focalplane.  Like the CCD data base, it is tagged.
	
	7. The ObsState data base defines the environmental state of an observation and
	the status of its subsequent photometry reduction. 
	
	8. The SodObs data base contains the photometry reduced by SoDoPhot for a single
	Tile's stars.
	
	9. The Tile data base defines the Tile boundaries. Although the information is
	derivable from data within the Domain data base, it is expanded 
	in order to speed photometry reduction and analysis.
	
	10. The TileHit data base defines every observation that overlaps a single Tile's 
	boundary. Although the information is available in the ObsState data base, 
	it is expanded in order to speed photometry reduction. 

\section{SYSTEM IMPLEMENTATION}

We now turn to some of the details of the actual system implementation.
The process-spanning Petri nets that are used to implement the Macho
system are based on a lower level set of interprocess communication
mechanisms.  These are discussed in Section 3.1.  The Petri nets
themselves are implemented as C++ classes that utilize the Sun LWP
library (in the SunOS implementation) or the Sun Threads library (in
the Solaris implementation).  This is discussed in Section 3.2.  The
maintenance cycle, which implements the heartbeat monitoring function
discussed in Section 1, is described in Section 3.3.  Section 3.4
describes the functioning of alarms, while 3.5 covers the data base
implementation, and 3.6 the user interface.

\subsection{Interprocess Communication Mechanisms}

Interprocess communication is handled by three mechanisms (Fig. 6): 
mailboxes, work queues, and the bulletin board.  Each of these is a 
persistent object in the sense described in Section 1.3.1, and is implemented 
as a memory mapped file.  

\subsubsection{Mailboxes}

Each process has a single mailbox to which any other process in the 
system can send messages.  A message consists of a header used for 
routing purposes, followed by an arbitrarily long string of bytes.  The 
byte string can be interpreted by a process in whatever way it 
wishes, for example as a C structure.  Messages are used to request a 
service from a process, to notify a process that some condition has 
occurred, or to request status information from a process.  A process 
is notified by the mail delivery system whenever a message arrives.

\subsubsection{Work queues}

Work queues are similar to mailboxes but differ in that they are 
shared by a group of recipient processes instead of being exclusively 
owned by a single process.   They are used, as the name implies, to 
deliver work to a group of server processes.  The first to become free 
will remove the item from the work queue and perform the 
requested task.

\subsubsection{Bulletin board}

In addition to the pipeline style of information flow supported by 
mailboxes and work queues, it is very useful to have a globally 
shared data structure that can be used to contain information that is 
of general use.  The bulletin board serves this function.  An item can 
be posted to the bulletin board by any process, and read by any 
process.  Items are identified by name and have an associated time 
tag so that customer processes know how current the information is 
that they are getting.  Items can be of arbitrary length, and as with 
messages, are usually interpreted as a C structure by customer 
processes.

The bulletin board is used to post items such as the number of the 
next observation, the current state of the telescope, temperatures 
from a variety of sensors, and observation descriptors for all 
observations that currently inhabit the pipeline.  

\subsection{Petri net}

The Petri nets are built on two C++ base classes, Place and
Transition.  Specialized types of Transitions, the TimeoutTransition
and ConditionalTransition, are derived from the Transition class by
inheritance.  Additional container classes, PlaceSet, TransitionSet,
Net, and SubNet, are used to bundle together related groups of Places
and Transitions, to define complete Petri nets, and subsets of Petri nets, 
respectively.  The Tokens that mark the Places of the Net are in
fact pointers to Messages, those objects that can be mailed from
process to process via the mail system.  If a Token does not need to
have any associated information, as is frequently the case, a null
pointer is used.

Places come in three flavors, Normal, Input, and Output.  Input Places
are marked by the arrival of a Message from another process addressed
to them.  The marking of an Output Place causes a Message to be
delivered to some Input Place in another process.  Normal Places do not
involve interprocess communication.  Each Place has an associated FIFO queue
for Tokens.  This modification to the basic Petri net behavior provides
for some buffering between Transitions, and improves performance.

The parallelism of the Petri nets is achieved through the use of the
Solaris Threads library \markcite{(SunSoft 1994)}, which allows applications to make use of
large numbers of parallel threads of control with low resource expenditure.
Each Transition has an associated thread, which implements the Petri
net firing rules and invokes any associated function when the firing
rules are satisfied.  At any moment the majority of these threads 
are suspended, waiting for the input Places associated with their 
Transition to be marked.

Associated with each Net is a single thread which processes Messages
arriving in the Net's Mailbox, and then distributes each one to the
Place for which it is bound.  The arrival of such a Message marks the
associated Input Place, and therefore may cause the associated Transition
to fire.  Similarly, marking of an Output Place spawns a thread 
which is responsible for sending the mail that will the deliver the
associated Message to its destination Input Place.

\subsection{Maintenance cycle}

In addition to the observation cycle, all processes participate in the 
maintenance cycle.  This cycle, which executes in parallel with the 
observation cycle, implements the heartbeat monitoring function.  
Any process can monitor the health of another by periodically 
sending a ``status request'' message to its mailbox.  If a ``status reply'' 
message is not received within a fixed timeout period, the 
heartbeat of that process is flagged as stopped.  The ``status reply'' 
message also can contain detailed information  about the internal 
state of the process.  Currently, both the UI and the Controller 
processes perform heartbeat monitoring functions.

\subsection{Alarms}

Any process that encounters an error condition, or any condition that
may require human intervantion, is free to post an alarm.  Alarms have
two types:  a ``Show'' alarm merely informs the operator, without requiring
a response, while a ``Resolve'' alarm requires an operator response.  As
examples of the two types, a Show alarm is sent on software startup to
inform the observer that the telescope is disconnected and must be
reconnected if he wishes to take a new field, while a Resolve alarm is sent
if disk space runs out and a file cannot be created.   The latter requires
an operator response because an operation attempted by the software (file
create) has failed, and further progress cannot be made until the
cause is corrected.

The response to a Resolve alarm is made by picking one of three choices
presented by the UI.  These are ``Retry'', ``Continue'', or ``Exit''.
Retry indicates the process should retry the failed operation, usually
after the operator has intervened in some way (e.g. freed some disk
space).  Continue indicates the process should simply continue
processing without attempt to handle the error, while Exit indicates
the process should terminate itself.  This latter option, which is
very rarely used, will then lead to another alarm when the Controller
realizes that the process has died, and thereby an opportunity to
start a new copy of the process.

Alarms are implemented using the mail system.  When a process posts an
alarm, mail is sent to the UI containing the text of an alarm message
and the identity of the poster.  If the type of the alarm is Resolve 
mail containing the choice made by the operator is returned to the 
poster, who then proceeds accordingly.

\subsection{Database}

In order to implement the Macho data base subsystem in minimal time,
we initially sought a commercial product. We 
quickly restricted our search to object oriented data base products, which 
more readily accomodate time history data ingest, manipulation, and retrieval
rather than do relational data bases.
We found the few existing (circa 1991) object oriented data base products 
to be too ``leading edge'' for production use.  We then reluctantly embarked 
on the development of a handcrafted data base system, the architecture of
which was sketched in Section 2.5.

The data base is not a monolithic object, but rather a large collection
of data base files organized by a multi-level directory tree.  The
majority of the data is maintained in specially constructed files which
are tuned for the data content and its ingest and retrieval patterns.
The data bases are implemented using a C++ library based on GNU C++
Associative Maps \markcite{(Lea 1992)}.  The base data base class
provides functions common to all data base types (open, close, create,
etc).  It also defines alternate access methods - either random disk
access or direct memory mapping.

Each specific data base class additionally defines the ingest, retrieval and 
status functions particular to the data.   For example, the segmentation and 
chaining of the time-sequenced photometry data base files is derived
from the base class and is tuned for efficiency of photometry installation 
and extraction. The data base classes, which provide a uniform 
interface, greatly simplify implementation of new data base types.

The Database system now manages more than 90,000 files, a number expected 
to grow to more than 200,000 as data collection continues.

\subsection{User Interface}

The UI receives and processes information from the pipeline in two ways. 
Firstly, it polls the bulletin board at regular intervals (normally every 5 
seconds) using XView notifier timer functions. This allows such items as the 
status of all processes and the contents of the pipeline (number and state of 
observation descriptors) to be updated on displays. Secondly, the UI sends 
and receives mail messages. These messages contain incoming alarms which are 
posted for action by the observer or outgoing requests for action by the 
controller or other component in the pipeline. Outgoing mail could be either 
observer responses to alarm conditions or intervention action like the 
removal of an invalid observation descriptor from the pipeline. 

On startup, the UI is presented as a root window with icons for each pipeline 
component. The icons are panel buttons that reveal customised sub-windows. 
Graphics, colour and animation are used widely in all sub-windows to 
enhance the observers ability to quickly assess the state of a process or to 
be alerted if action is required.  

The UI is written in C and uses the XView 3.0 toolkit \markcite{(Heller 1991)}.

\section{EXPERIENCE WITH THE SYSTEM}

We have now had over two years of experience with the system described
here, during which over 30000 observations have been taken and nearly 3 terabytes of
data collected.  For the most part, it has fully met our expectations and continues
to fulfill the tasks originally set for it.  The best features of the
system have been:

	1. Rapid implementation
	
	Design of the system began in July of 1991.  Stable operation of the
	first version of the system was achieved in September of 1992 with
	an expenditure of about 2.5 man-years.  This version lacked many
	capabilities that were added later, such as flatfielding of images
	and online photometry, but it was sufficient to collect data for
	analysis.  
	
	2. Ease of modification

	The system has been continually modified since it first became
	operational, both in incremental, and occasionally in more major
	ways.  The incremental modifications have generally been to add
	new features or correct bugs, while the major changes have been
	associated with changes in the underlying hardware and OS support.
	For the most part, these changes have been surprisingly easy to
	make.  Most importantly, none of them have required significant
	change to the underlying Petri net and data base concepts that the system is
	based on.  This stability, combined with the extensibility of
	C++, has made the system easy to live with in the real world.
	
	3. Robustness

	The system has turned out to be quite robust to a variety of
	failures, as we originally intended.  The original
	multiprocessor it was implemented on suffered many system
	crashes, and the use of persistent state in the observing
	software allowed it to be restarted after a system crash with
	minimal loss of time or data.  Similarly, the system recovers
	well from crashes caused by its own bugs, with the ability to
	easily restart a component process generally making recovery
	fairly painless. The use of alarms has also turned out well,
	providing a simple but flexible way for the operator to
	usefully respond to the exceptional conditions that always
	arise in the real world.

	4. Performance

	The performance of the system has increased continuously, as
	more powerful hardware has been added and software has been
	improved.  Today the pipeline meets all of the performance
	goals that we set for it.  First, the interval between
	successive images is within a few seconds of the minimum set by
	exposure time + ccd readout time.  This maximizes the rate at
	which raw data is taken.  
	
	Secondly, photometric reductions and data base loading happen
	nearly as quickly as images are taken.  This has been critical
	to the project's ability to ``alert'' on microlensing events
	very shortly after they begin.  Currently, data from a full
	winter night's observing that ends at 6 am are typically fully
	processed by 8 am when using 4 cpu's for photometry.  Templates
	have been made for only about half of our fields, however.
	Although our current templates include all of the densest
	fields, and therefore represent more than half of the total
	stars, a full set of templates will significantly increase the
	processing load.  When we have the full set, we will need to
	use 7 or 8 cpu's for photometry (out of the 12 available) to
	maintain the same performance.

	Finally, data base retrieve performance is good, and adequately
	supports analysis of lightcurves.

The problem areas have been:

	1. Bifurcation of the system

	As stated above, the initial version of the system supported
	neither flatfielding nor photometry.  With data pouring in,
	there was a powerful motivation to complete the analysis path
	as quickly as possible.  There was a division of opinion as to
	whether this should be done by completing the data pipeline, or
	by creating a separate ad-hoc ``offline'' system fed from the
	existing ``online'' data pipe to perform the analysis.  In the
	event, the development of the system bifurcated, with an
	offline system being created while data pipeline development
	continued at a slowed rate.

	Today, both systems are complete and are in active use.  The
	offline system performed the photometry and analysis used in
	all of the project's publications to date.  The online system
	is currently used for all new photometric reductions, is the
	only repository for recent photometry, and provides the only
	capability for alerts.  Older photometric data is partially
	duplicated between the two systems, with the online system
	expected to have a complete set of the project's photometry
	within the next six months.

	Although our plan is that the online system will ultimately
	be the basis for all the project's analysis, this is not an
	easy transition to make.  Each is the result of a large effort,
	and has a group of committed users.  It is likely that we
	will be dealing with the effects of the bifurcation for
	the life of the project.

	2. Difficulty of text-based Petri net specification

	Petri nets are most naturally specified and thought about
	graphically.  Our software, however, requires a textual
	specification of the net which is directly input to the C++
	compiler.  This has caused a continual, and costly, mismatch
	between the human specification and the machine specification.
	It has been the most fertile source of bugs, which have 
	frequently arisen because the text specification was an
	improper translation of the graphical object that the designer
	had in mind.
	
	3. Difficulty of debugging
	
	Debugging parallel, real-time applications is difficult at the
	best of times.  In addition to this generic problem, some
	specific difficulties made our lives difficult.  One of these,
	of course, is the problem of rapidly and reliably translating
	back and forth from the text-based Petri net specification
	required by the compilers to the graphical notation required by
	human understanding, as mentioned above.  Additionally, most of
	the life of the system has been under SunOS 4.1, where the
	Petri nets were built on top of the LWP library.  This library
	had many bugs and little support, and therefore required much
	investment of time to find workarounds.  This problem has gone
	away with the switch to Solaris and the Threads library.
	Similarly, under LWP there was no debugger support of any kind,
	which made it necessary to debug largely with print statements!
	The debugger under Threads has sophisticated built-in support
	for threads, and is making life much easier.

	4. Database issues
	
	Although the hand-crafted data base system is efficient, robust,
	and reliable, its greatest strength  - being a no frills data
	entry and retrieval engine - is also its greatest liability.
	Database record definitions are difficult to alter. Definition
	changes require either retrofitting the extant data base records
	or using versioning to determine, on the fly, the appropriate
	record format to use. To date, we have opted for the former.

	An early decision was taken to physically load photometry
	records in time-ordered sequence in the anticipation that this
	would greatly enhance data retrieval speed. This subverted the
	classic data base management premise that a record is extracted
	by its selection key, and has caused a number of difficulties.
	In the event, all our current analysis tools access records via
	key.

	5. User interface and XView
	
	The high-level nature of the XView calls (as compared to raw
	Xlib or Motif) allowed the UI to be developed quickly.
	However, XView has proven to be a poor choice for
	a UI that relies on timers and interrupts. Incoming mail in the
	SunOS implementation of the pipeline was signalled to the UI
	using the standard UNIX signal SIGUSR1. This required an
	asynchronous interrupt to the notifier loop that often failed
	and caused the UI to crash.
	Furthermore, this situation is exacerbated by the presence of
	multiple notifier timer functions that periodically read the
	bulletin board and run animation graphics. XView under Solaris
	seems to have similar notifier interaction problems. A robust
	UI in an interrupt driven and polling application like the
	pipeline will probably require a move to a toolkit better
	suited to realtime applications.

	6. Stress on system software

	We rapidly discovered that this system works many features
	of the operating system software quite hard.  In addition to
	the persistent LWP problems mentioned above, there were 
	many early problems due to our intensive use of memory
	mapping and network locking of memory mapped files.  Unreliable
	signal delivery has also been a major problem.  These
	problems have generally gotten better, but finding workarounds
	has been time consuming.

\section{FUTURE DEVELOPMENT}

Looking to the future, we expect to operate the system in a fully
distributed mode, where member processes can exist on fully separate
machines (today they can run on multiple processors, but all on the
same machine).  Although the underlying process-spanning Petri net
model fits perfectly into a fully distributed environment, the 
low level machinery will not make the transition readily.  Largely,
this is because the Solaris Threads library itself is not designed
to cross machine boundaries.  We have not yet decided how best to
take this next step.

Additionally, we expect to implement a graphical method for
specifying the Petri nets that form the core of the system.  This 
software is straightforward (although time consuming) to create, and
will be a major aid to system understanding and debugging.  Also, it will
help ensure that documentation and software are kept synchronized.
As part of this effort, we will reexamine the features of
our extended Petri nets, and possibly modify them to reduce the need
to propagate state information outside of the nets.

There are a number of ways in which we plan to improve and extend
the pipeline.  The two most important are to include a process which
analyzes images in the pipeline to automatically focus the telescope;
and to include the ``alert'' software, which detects new microlensing events,
within the pipeline.

Finally, the volume of our reduced data is rapidly outstripping our
ability to purchase online disk storage for it.  We plan to move to a
system that holds older data on ``nearline'' storage at the ANU
Supercomputer Facility (ANUSF), transparently migrating it as it is
required.  Our data base organization, based as it is on large numbers
of individual files, will adapt well to this form of distribution.

\section{FIGURES}

\begin{figure}
\plotone{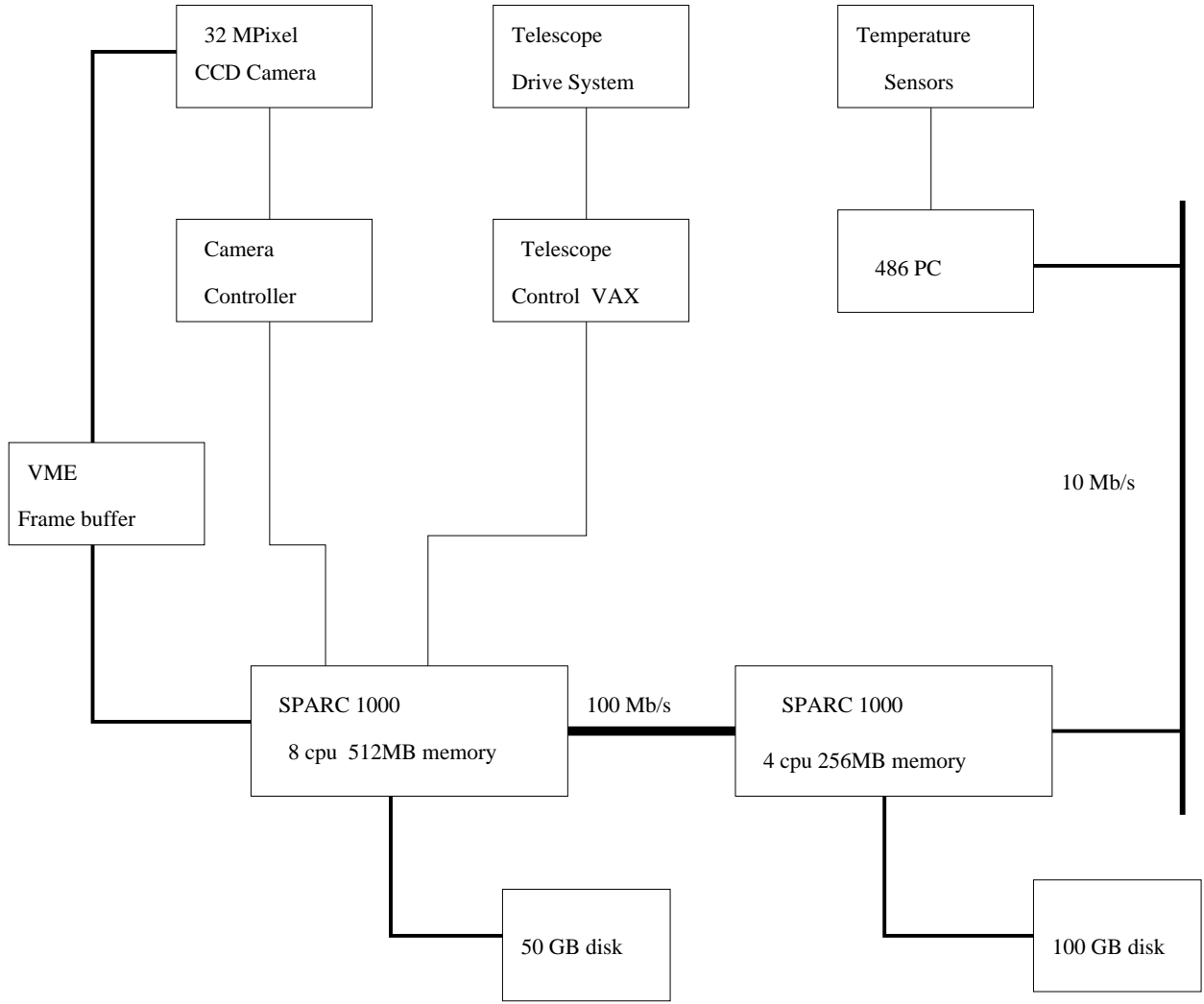}
\caption{Major components of the MACHO computer hardware}
\end{figure}

\begin{figure}
\plotone{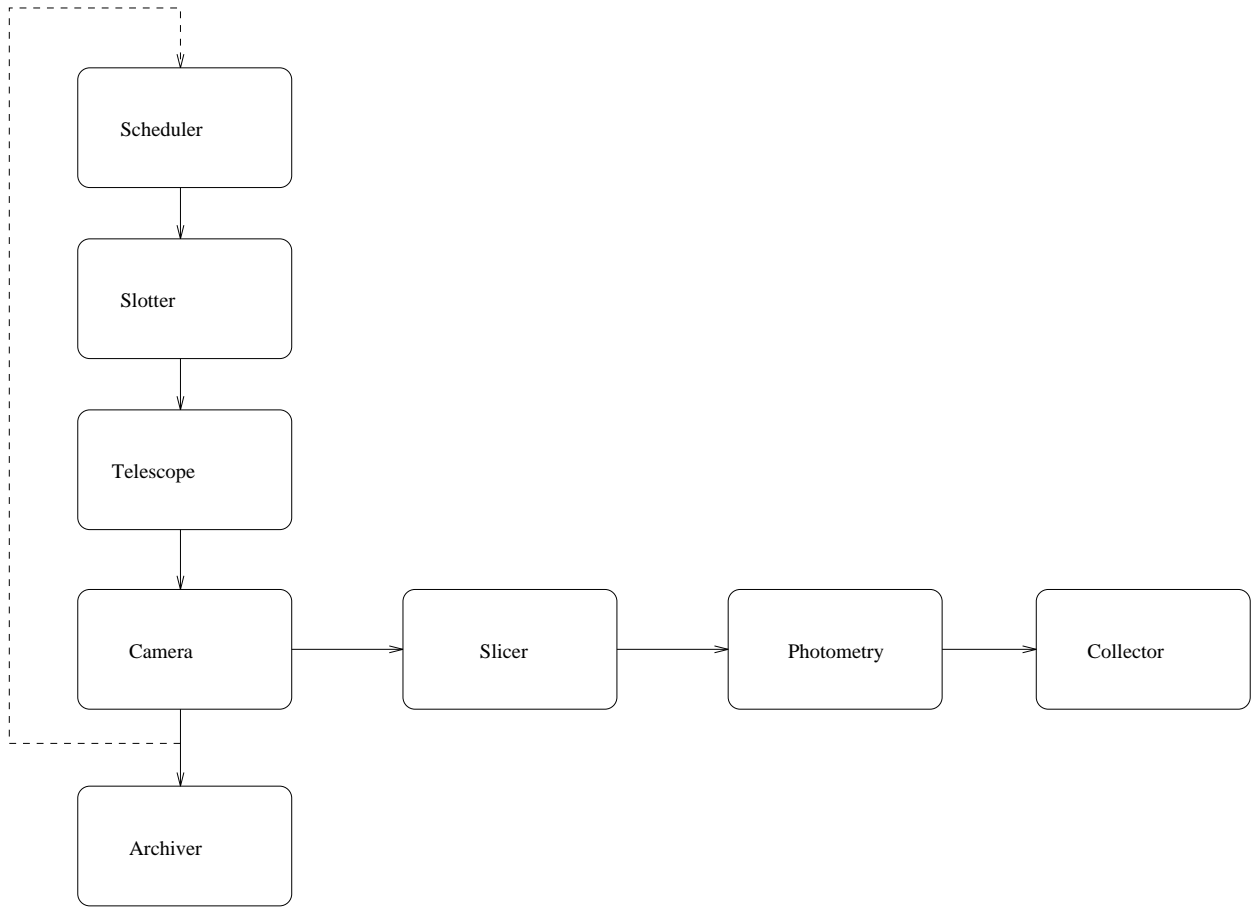}
\caption{The processes that make up the observation cycle are shown.  The arrows connecting them indicate the sequence in which the operations are performed for each observation.}
\end{figure}

\begin{figure}
\plotone{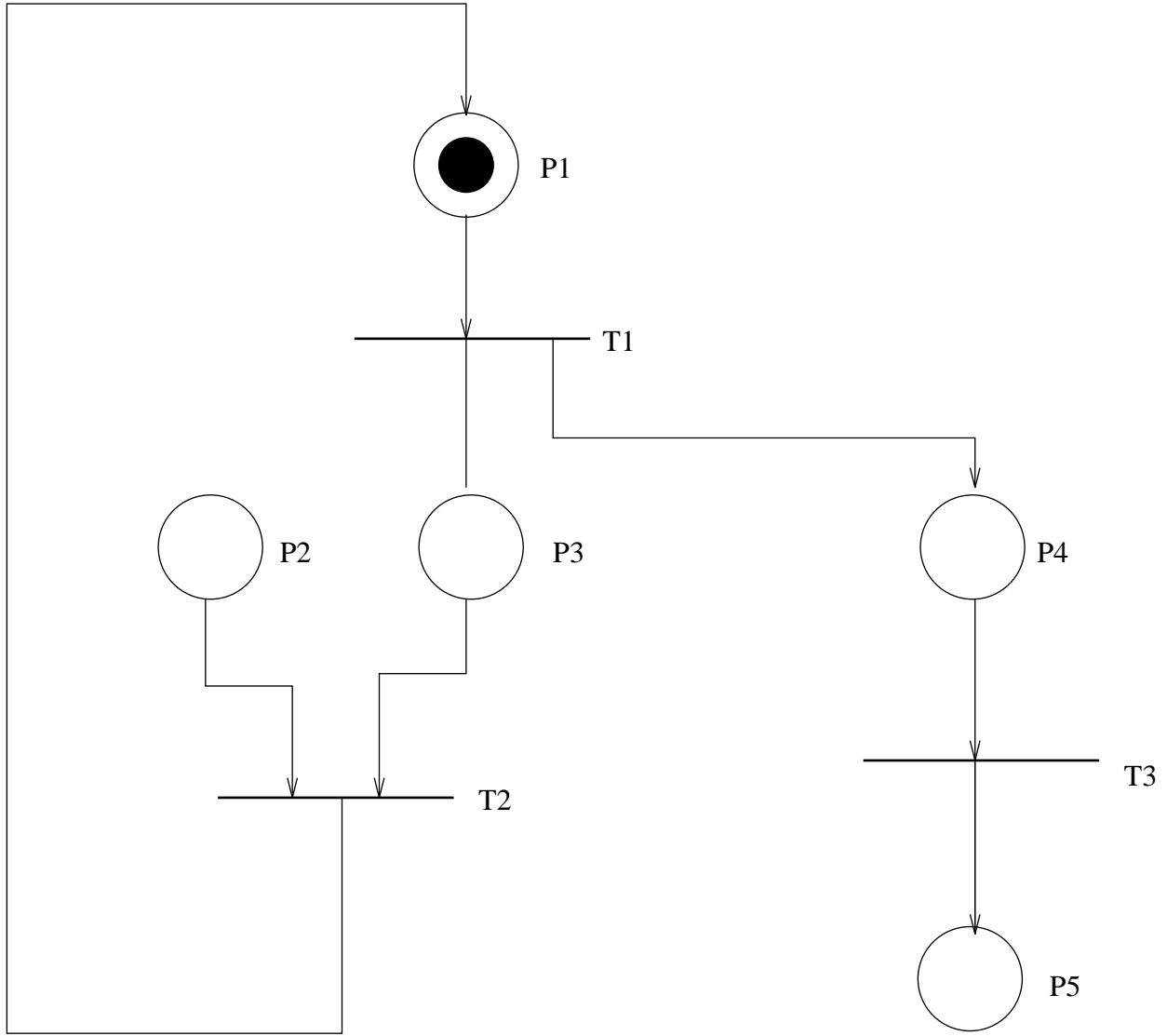}
\caption{A simple example of a Petri net.  The circles denote places and the horizontal lines transitions.  The filled circle in place P1 indicates it is marked with a token.  This will allow transition T1 to fire, thereby marking places P3 and P4.  T3 will then fire, marking P5.  Unless place P2 is marked, no further activity can take place in this net.  If it is marked by some external agency, however, Transition T2 will then fire, and the whole cycle will begin again.}
\end{figure}

\begin{figure}
\plotone{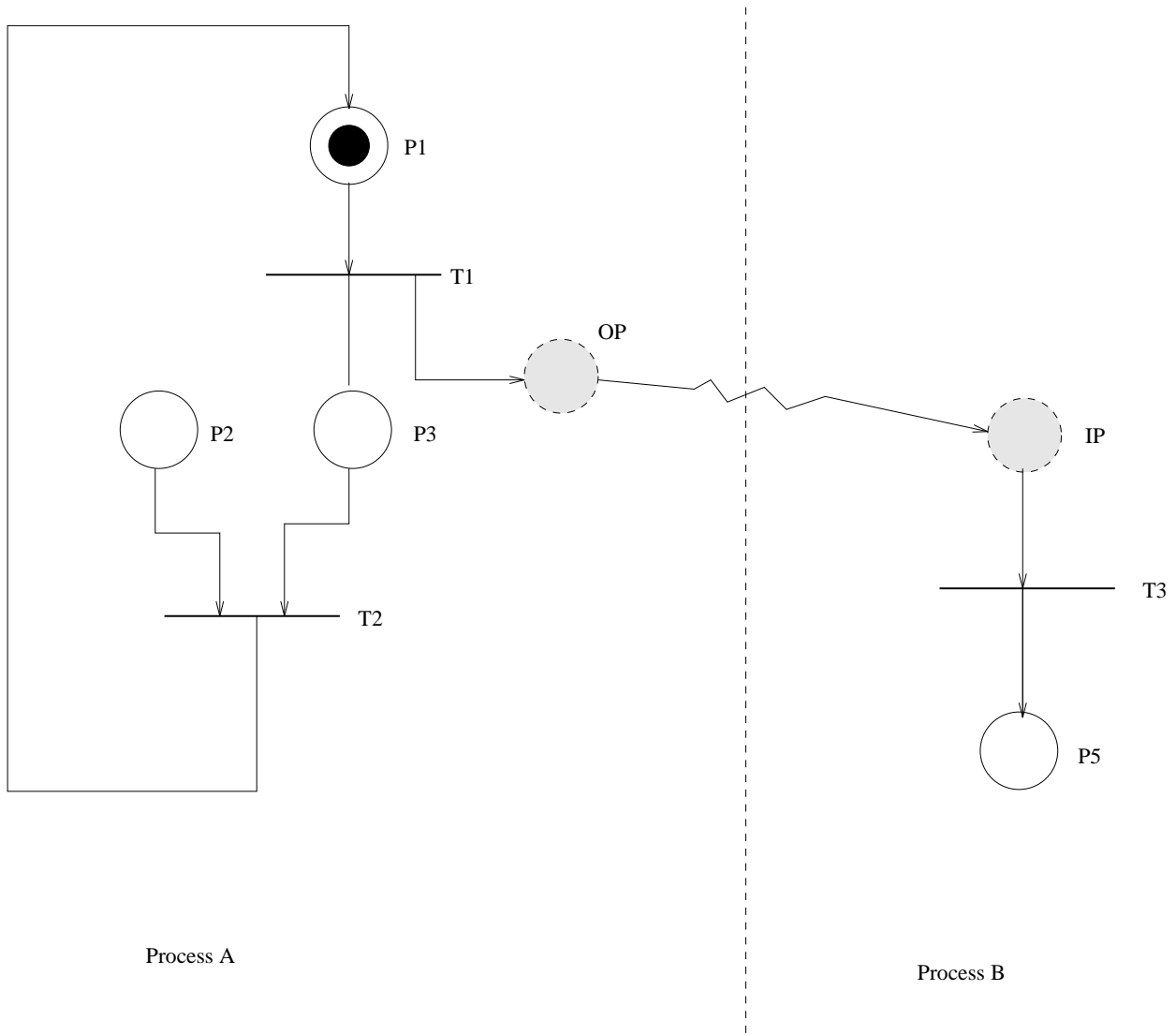}
\caption{The same net as in Fig. 3 is shown, but now the dotted line shows the boundaries between two separate processes, A and B.  When transition T1 fires, output place OP is marked, causing an interprocess message to be sent that marks the input place IP in process B. The functionality of the net is exactly the same as before.}
\end{figure}

\begin{figure}
\plotone{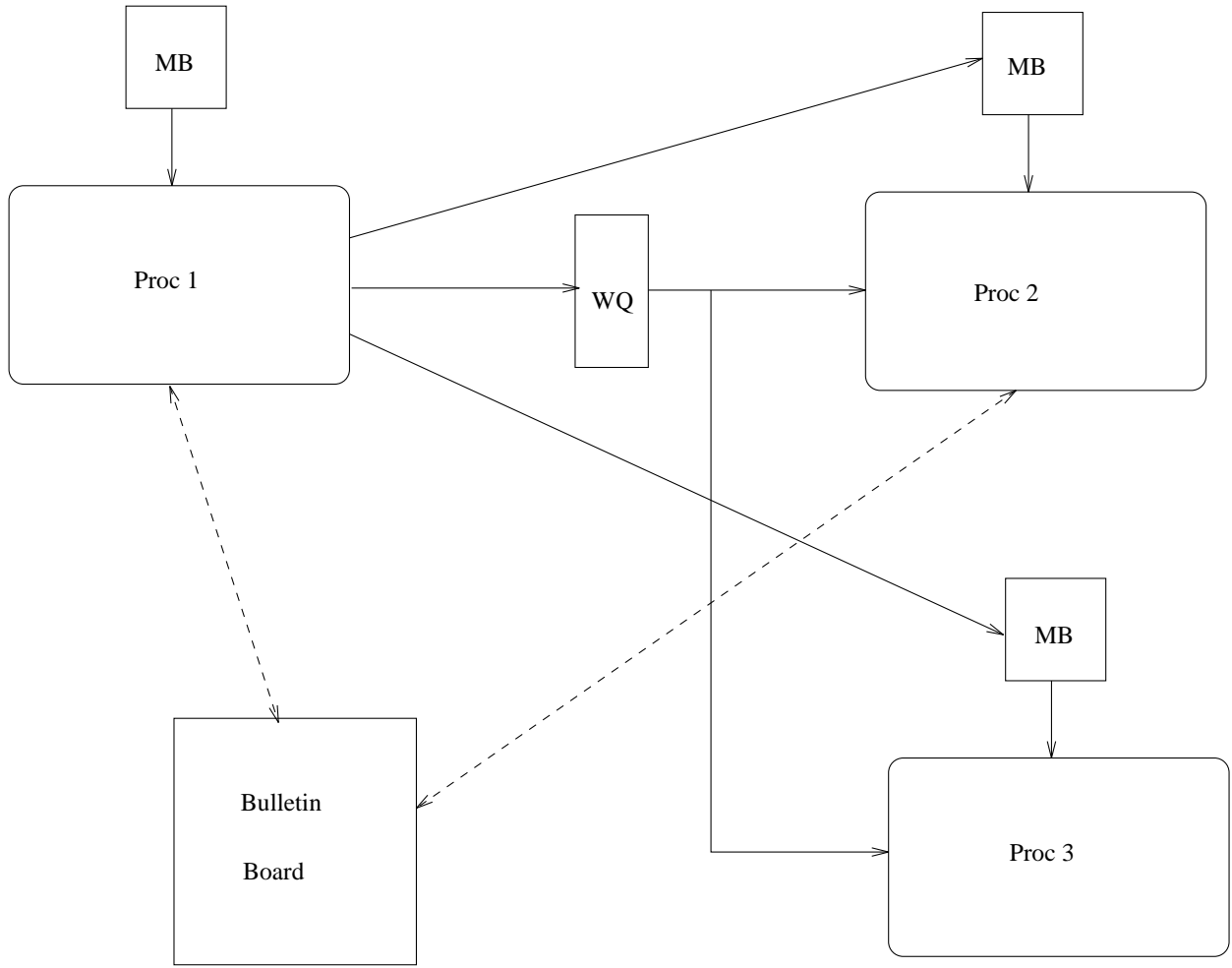}
\caption{Geometry of fundamental data base structures}
\end{figure}

\begin{figure}
\plotone{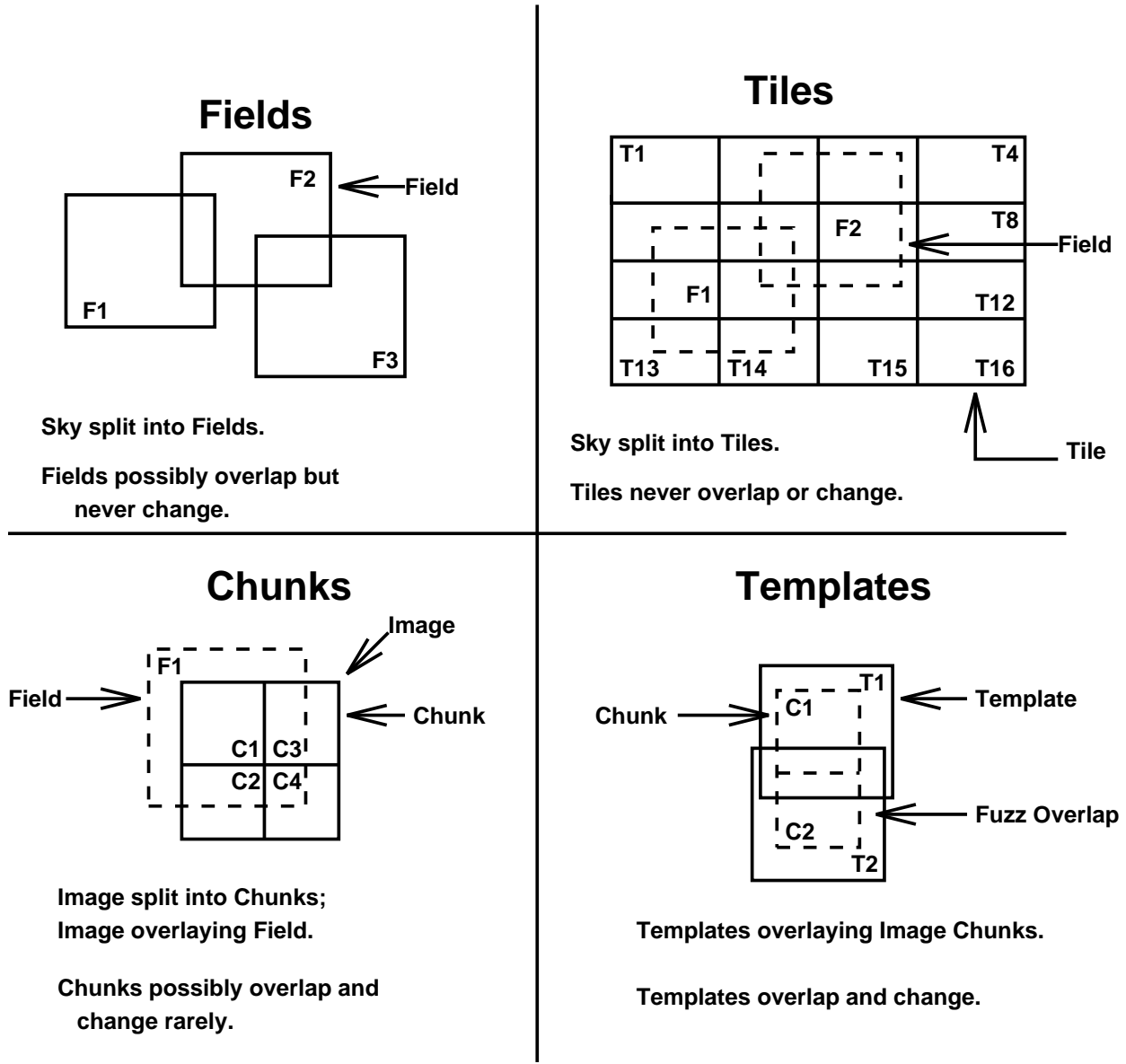}
\caption{The functionality of the various interprocess communication entities are indicated.  MB denotes a mailbox, and WQ a work queue.  Note that processes 2 and 3 share the same work queue.  A work queue item will be processed by whichever of them is first free.}
\end{figure}

\newpage
\begin{verbatim}

Table 1.  Display of the information contained in a typical observation descriptor.

observation state: 
     obs id: 33415 
     ObsNumber: 33415     	time: Wed Jul 26 18:25:26 1995 
     	field_id: 113	    RA: 18:0:20.9363  DEC: -28:56:45.12 
     	telescope pointing: RA: 18:00:20.8514 DEC: -28:56:45.8466  West of Pier 
     	exposure time: 150 
     	seeing: 0 
     	airmass: 1.37137 
     	parallactic angle: 4.32645 
     	refraction: 58.2 
     	focus: 3.92 
     Locators:  
     	Active: /meta_tmp/PActive 
     	Archive: /meta_tmp/PArchive 
     	Camera Configuration: test.camconfig 
     	archive image CCD tag: 950723 
     Camera Status: 
     	readout[0]: 1100 
     	readout[1]: 2200 
     	readout[2]: 0 
     	readout[3]: 0 
     	readout[4]: 1124 
     	readout[5]: 2148 
     	readout[6]: 1 
     	readout[7]: 1 
     	readout[8]: 0 
     	readout[9]: 0 
     	readout[10]: 0 
     	readout[11]: 0 
     	readout[12]: 10 
     	nclears1: 0	nclears2: 5	nwait: -1 
     Controller flags: 
     	check_list: 0x101ff 
     	obs_type: 0 
     	reobs_cam_tag: 0 
     	reobs_date: invalid     	reobs_check_list: 0x0 
     	photometry type: Sod 
     Flatfielding status: 
     	flags: XTALK BYROW DARK VOODOO1 FLAT RESCALE STATS WARN 
     	rescaled_above	= 62200 
     	config_file_tag	= 3 
     	ctm_file_tag	= 4 
     	dark_image_tag	= 5 
     	flat_image_tag	= 7 
     	debias_region	= [1071,1120:1,2148] 
     	voodoo1_rows	= 701 to 1041 
     	amp_info: 
     	             gain    itop   bias   noise     sky  xtalk   dark  stuck 
     	    _0_0:   0.827   78650      0    4.14       0      2      0   FF80 
     	    _0_1:    1.45   57475      0    4.40    2687      3     -6   0000 
     	    _1_0:    1.59   57475      0    3.97    2626      4     -2   0000 
     	    _1_1:    1.48   64735      0    4.24    2670      0      0   0000 
     	    _2_0:    1.47   57475      0    4.36    2523      4     -1   0000 
     	    _2_1:    1.52   58685      0    4.09    2508     -2      0   0000 
     	    _3_0:    1.54   59895      0    4.03    2176      0     -1   0000 
     	    _3_1:    1.52   58685      0    3.81    2592      3     -2   0000 
     	    _4_0:    1.66   61105      0    4.36    1450     -1      1   C000 
     	    _4_1:    1.53   58685      0    3.89    1600      2      1   0000 
     	    _5_0:    1.52   58685      0    4.84    1803      3      0   8000 
     	    _5_1:    1.67   53845      0    5.38    1600      0      0   8000 
     	    _6_0:    1.50   50215      0    5.23    1660      0      0   0000 
     	    _6_1:    1.42   50215      0    5.19    1701      0      0   8000 
     	    _7_0:    1.40   51425      0    4.51    1721     -1      0   0000 
     	    _7_1:    1.33   50215      0    4.55    1794      1      0   8000 
     Sensor log: 
     	time_stamp: Wed Jul 26 18:24:56 1995 
     	      Tout = 5.084 C 
     	     Tinsd = 5.815 C 
     	     Tmtlw = 5.645 C 
     	     Ttlwr = 5.385 C 
     	     Tmirr = 6.122 C 
     	     Tmtmd = 5.690 C 
     	     Ttmid = 5.449 C 
     	     Ttupr = 6.261 C 
 
     Dap template tag: -99999999 
         PR source version: -99999999 
         PR CCD tag: -99999999 
         PR date: invalid 
     Sod group template tag: 940510 
         PR source version: 87 
         PR CCD tag: 950723 
         PR date: Wed Jul 26 18:34:33 1995 


     chunk count: 121 
     chunk: 0 
     chunk: 
          chunk id: 0 
          sod template observation: 7793 
                       star count:  9582 
              psf parameter(0): 9.273 
              psf parameter(1): 3.021 
              psf parameter(2): 2.708 
              psf parameter(3): -0.03729 
              average sky value(0): 9337 
              average sky value(1): 3986 
              transformation parameter(0): 1.003 
              transformation parameter(1): 0.000962 
              transformation parameter(2): -0.0009312 
              transformation parameter(3): 1.003 
              transformation parameter(4): -16.3 
              transformation parameter(5): 6.021 
              transformation parameter(6): 0.0109 
              transformation parameter(7): -0.01907 
              transformation parameter(8): 0 
              transformation parameter(9): 0 
              transformation parameter(10): 0 
              transformation parameter(11): 0 
          dap template observation: 0 
          ddap template sky per pixel: 0 
               sky per pixel: 0 
               aperture size used: 0 
               sky correction: 0 
               flux calculation: 0 
               transformation parameter(0): 0 
               transformation parameter(1): 0 
               transformation parameter(2): 0 
          # normalization stars: 25 
          normalization stars 
               star id:: tile: 18942   seqn: 42 
               star id:: tile: 19072   seqn: 47 
               star id:: tile: 18942   seqn: 44 
               star id:: tile: 18942   seqn: 46 
               star id:: tile: 18941   seqn: 50 
               star id:: tile: 18941   seqn: 53 
               star id:: tile: 19071   seqn: 57 
               star id:: tile: 18941   seqn: 54 
               star id:: tile: 18941   seqn: 59 
               star id:: tile: 18941   seqn: 60 
               star id:: tile: 18942   seqn: 53 
               star id:: tile: 18942   seqn: 54 
               star id:: tile: 18942   seqn: 55 
               star id:: tile: 18942   seqn: 56 
               star id:: tile: 18943   seqn: 20 
               star id:: tile: 18941   seqn: 69 
               star id:: tile: 18942   seqn: 62 
               star id:: tile: 19072   seqn: 66 
               star id:: tile: 18942   seqn: 64 
               star id:: tile: 18942   seqn: 65 
               star id:: tile: 18943   seqn: 21 
               star id:: tile: 18942   seqn: 67 
               star id:: tile: 18941   seqn: 74 
               star id:: tile: 19071   seqn: 76 
               star id:: tile: 18941   seqn: 75 

..... 
Information from additional 120 chunks deleted 
..... 

     byte count of Sod input parameters: 1878 
     Sod Input Parameters: 
N0STARMN = 6 
SEEMIN = 2.0 
HWHM_TOL = 2.0 
NPIXCELL = 10 
AXIS_RATIO = 1.0 
TILT = 0.0 
ISKYROWMN = 25 
ISKYROWMX = 500 
ISKYROWINT = 25 
SKYPERCENT = 0.3 
DELSKYHIST = 30. 
SKYMIN = 0. 
SKYMAX = 40000. 
TOLMATCH = 3. 
NFITBOX_X = 11 
NFITBOX_Y = 11 
MASKBOX_X = 5 
MASKBOX_Y = 5 
APBOX_X = 17.0 
APBOX_Y = 17.0 
THRESHMIN = 40.0 
THRESHMAX = 20000.0 
THRESHDEC = 1.0 
THRESHDEC2 = 1.5 
THRESHDEC3 = 1.5 
NLOOPMN = 2 
NLOOPMN2 = 2 
NLOOPMN3 = 2 
IDTYPE = 1 
LEVPSFFIT = 2 
ISAMEC = 1 
VMINUSR = 0. 
DFMAGSIG = 5.0 
DIFMAGMN = 0.05 
FWHMPSF = 5.0 
ILOOKUP = 1 
NPERFWHM = 80 
STPAR2CRIT = 20000. 
NSTRFIND = 100 
NTRI_IMG = 5000 
NSTAR_FID = 100 
NTRI_FID = 5000 
NTRI_MATCH = 100 
MIN_SET_SIZE = 35 
BOX_RADIUS = 1 
NTCELL = 40 
DMIN = 30. 
DMAX = 700. 
TOL = 2.0 
SLOPE = 0.005 
FAT = 20. 
HIST_LEVEL = 0.003 
MN_MATCH = 7 
DEL_X_MX = 275. 
DEL_Y_MX = 275. 
THETA_MX = 0.03 
DEL_X_TOL = 7.0 
DEL_Y_TOL = 7.0 
THETA_TOL = 0.015 
R_TOL = 2.0 
RATIN_MX = 1.10 
COLORTERM_V = 0.108 
COLORTERM_R = 0.025 
MAGFIDNORM = 0 
SCALE_RAT0 = 1. 
NPOSTFIT = 1 
NSKYBOXS2 = 8 
ROLDMX = 2.5 
PRETHRESHFAC = 3. 
CHI2CUT = 10000. 
DMAGCUT = 0.1 
DMSECUT = 7.0 
RXFITMX = 5.0 
NSKYAVX = 10 
NSKYAVY = 10 
NSKYGRIDX = 40 
NSKYGRIDY = 40 
AUTOTHRESH = 'YES' 
OBJ_IN_NORM = 1. 
LOGVERBOSITY = 1 
FOOTPRINT_NOISE = 1.3 
NPHSUB = 1 
NPHOB = 1 
ICRIT = 40 
CENTINTMAX = 2.5e5 
CTPERSAT = 8000. 
STARGALKNOB = 8.e6 
STARCOSKNOB = 1.0 
KCOS = 2 
SNLIM7 = 7.0 
SNLIM = 0.5 
SNLIMMASK = 4.0 
SNLIMCOS = 3.0 
NBADLEFT = 0 
NBADRIGHT = 0 
NBADTOP = 0 
NBADBOT = 0 
SKYTYPE = 'PLANE' 
OBJTYPE_IN = 'SMALL' 
OBJTYPE_OUT = 'UNFORM' 
BADTYPE_IN= = Input 
IALERT = 1 
NFITBOXFIRST_X = 31 
NFITBOXFIRST_Y = 31 
CHI2MINBIG = 16 
XTRA = 25 
SIGMA1 = 0.10 
SIGMA2 = 0.10 
SIGMA3 = 0.10 
ENUFF4 = 0.50 
ENUFF7 = 0.80 
COSOBLSIZE = 0.9 
APMAG_MAXERR = 0.1 
PIXTHRESH = 1.0 
BETA4 = 1.0 
BETA6 = 0.5 
FITBOXMIN = 5.0 
SCALEAPBOX = 6.0 
APBOXMIN = 7.0 
SCALEMASKBOX = 1.5 
AMASKBOXMIN = 5.0 
SIGMAIBOTTOM = 10.0 
SIGMATHRESHMIN = 2.0 
END = = 
\end{verbatim}

\end{document}